\documentclass[11pt,twoside]{article}
\usepackage{asp2004}
\usepackage{epsf}
\usepackage{graphics}
\usepackage{lscape}
\usepackage{verbatim}
\usepackage{natbib}
\def\edcomment#1{\iffalse\marginpar{\raggedright\sl#1\/}\else\relax\fi}
\reversemarginpar

\begin{document}
\title{The first PG\,1159 close binary system}
\author{S.\ Schuh}
\affil{Institut f\"ur Astrophysik, Universit\"at G\"ottingen,
  Friedrich-Hund-Platz~1, D-37077~G\"ottingen, Germany}
\author{T.\ Nagel}
\affil{Institut f\"ur Astronomie und Astrophysik, Universit\"at T\"ubingen,
  Sand~1, D-72076~T\"ubingen, Germany}

\begin{abstract}
The archival spectrum of \mbox{SDSS~J212531.92-010745.9} shows not only
the typical signature of a PG\,1159 star, but also indicates the
presence of a companion. With time-series photometry of
\mbox{SDSS~J212531.92-010745.9} during 10 nights, spread over one month, with the
T\"ubingen 80\,cm and the G\"ottingen\,50 cm telescopes, the 
binary nature of this object has recently been proven.
An orbital period of 6.9\,h could be determined, and the
observed light curve fitted with the \texttt{nightfall} program. 
A comparison of the spectrum of \mbox{SDSS~J212531.92-010745.9} with NLTE
models further constrained the light curve solution. We emphasize that
this is the first system of this kind which will allow a dynamical mass
determination for a PG\,1159 star.
\end{abstract}
\section{Masses of PG\,1159 stars}
PG\,1159 stars are hot hydrogen-deficient pre-WDs which are believed
to be the outcome of a late (or very late) helium-shell flash. 
Their non-standard evolutionary history, including a second passing
through post-AGB stages, is reflected in their unusual surface
composition (see also \citealt{reiff:07,jahn:07}) 
which is characterized by strong hydrogen-depletion.
Spectroscopic masses have in the past usually been derived from
hydrogen-rich models, using the sets by 
\citet{1983ApJ...272..708S} and 
\citet{1995A&A...299..755B} and
\citet{1986ApJ...307..659W}.
\subsection{Spectroscopic masses}
The derivation of spectroscopic masses requires to obtain accurate
photospheric parameters ($T_{\rm eff}$, $\log\,{g}$) and combine them
with evolutionary model calulations. The most comprehensive
compilation, using NLTE stellar atmosphere on the one hand and actual
late helium-shell flash modelling on the other hand, has
recently been presented by \citet{2006PASP..118..183W}, collecting
together a total of 40 objects in their analysis. A further set of
state-of-the-art evolutionary tracks has simultaneously been presented by
\citet{2006A&A...449..313M}. Both finally make available
late thermal pulse post-AGB evolutionary models, but differ in
details on which mass track actually corresponds to a given position in
$T_{\rm eff}$, $\log\,{g}$ diagrams.
\subsection{Asteroseismic masses}
A complementary approach uses the fact that ten of the currently known
PG\,1159 stars lie in the GW~Vir instability strip (for the current
status of work trying to theoretically and empirically constrain the
red and blue edges of the GW~Vir instability strip, see the
contributions by \citealt{quirion:07,vauclair:07,werner:07}). 
From an analysis of the GW~Vir stars' pulsational eigenmodes,
fundamental parameters including the 
mass can be derived through comparison to structural models 
(usually requiring some additional spectroscopic information to narrow down
the number of possible solutions). Such solutions have been found for five
objects so far (Table~\ref{tabpulsators}).
\begin{table}[!t]
\caption{Results from asteroseismology of PG\,1159 stars and the [WC4]
central star of NGC~1501. This compilation (adopted from Table~3 of
\citealt{2006PASP..118..183W}) compares the stellar mass derived by
spectroscopic means M$_{\rm spec}$ with the pulsational mass M$_{\rm
puls}$. Other columns list envelope mass M$_{\rm env}$ (all masses in
solar units) and rotation period P$_{\rm rot}$ in days.}
\label{tabpulsators}
\smallskip
\begin{center}
{\small
\begin{tabular}{lcclll}
\tableline
\noalign{\smallskip}
Star            &M$_{\rm spec}$&M$_{\rm puls}$&M$_{\rm env}$&P$_{\rm
  rot}$&Reference \\
\noalign{\smallskip}
\tableline
\noalign{\smallskip}
PG\,2131+066    & 0.58         & 0.61         & 0.006       & 0.21 & \citet{kawaler:95} \\
PG\,0122+200    & 0.58         & 0.59         &             & 1.66 & Fu et al.\ (submitted) \\
RX\,J2117.1+3%412
                & 0.70         & 0.56         & 0.045       & 1.16 & \citet{vauclair:02a} \\
PG\,1159$-$035  & 0.60         & 0.59         & 0.004       & 1.38 & \citet{kawaler:94} \\
PG\,1707+427    & 0.59         & 0.57         &             &      & \citet{kawaler:04} \\
NGC 1501        &              & 0.55         &             & 1.17 & \citet{1996AJ....112.2699B} \\
\noalign{\smallskip}
\tableline\
\end{tabular}
}
\end{center}
\end{table}
\subsection{Dynamical masses}
Both of the above methods require a substantial amount of modelling,
incorporating assumptions on the evolutionary state of these objects
that one would ideally want to constraint with independently derived
measurements of the mass in the first place. 
About the only way to do this with as little a priori information as
possible is to dynamically weigh the components in a binary system.
\par
NGC~246 has long been known to be one component in a wide pair, but at
a separation of 3\hbox{$.\!\!^{\prime\prime}$}8 from its companion, an
observation of the dynamical aspect of their orbit is out of reach.
It is still very useful, though, as the distance to the system and
surrounding planetary nebula could be derived from photometry of the
companion \citep{1999PASP..111..217B}.  Furthermore, it has recently
been claimed to be pulsating (\citealt*{2006A&A...454..527G}).
\par
Yet another pulsator, PG\,2131+066, was found to show an 3.9\,h
periodicity, in addition to the shorter pulsational photometric
variations, by \citet*{1998A&A...331..162P}, who attributed this to
orbital motion in a close binary. The close binary hypothesis was
supported by the observation of a red
excess \citep*{1985ApJS...58..379W} and H$\alpha$ and H$\beta$ emission
superimposed on the PG\,1159 spectrum,
but was disputed by \citet*{2000ApJ...545..429R}, who measure a
physical companion 0\hbox{$.\!\!^{\prime\prime}$}3 away from the
primary in HST WFPC1 images and argue that the emission is caused by
that farther-away companion alone. The claim that PG\,2131+066 resides
in a close binary therefore remains unconfirmed.
\par
The only other PG\,1159 star known to be part of a binary system is
the recently discovered \mbox{SDSS\,J212531.92$-$010745.9}, and it is
finally without doubt also a close one, i.e.\ suitable for dynamical
mass determination.
\section{SDSS\,J212531.92$-$010745.9}
\citet{schreiber:07} systematically searched SDSS archival data for
WD+MS companion candidates and discovered H$\alpha$ emission and a red
colour excess in SDSS\,J212531.92$-$010745.9. The candidate was
further analysed by \citet{2006A&A...448L..25N}; we briefly
recapitulate their main results below.
\subsection{SDSS spectrum}
To fit a composite spectrum to the SDSS data, the model grid and
fitting method described by \citet{huegelmeyer:07} has been employed,
but a more complex iterative procedure was required to account for the
contribution of the secondary. As the latter was realized as a simple
black-body contribution, the best-fit composite ($T_{\rm eff} =
90\,000\,$K, $\log\,g\,[\rm cm\,s^{-2}] = 7.6$ for the PG\,1159
component) obviously fails to reproduce the emission line features.
In the PG\,1159 spectrum, \textup{C\,{\mdseries\textsc{iv}}} is
furthermore too weak in the model. This only reinforces, however, the
identification as a genuine PG\,1159 star, and prompted
\citet{2006A&A...448L..25N} to secure time-series photometry of the
object in the 2005 observing season.
\subsection{Light curve}
Photometric observations of \mbox{SDSS\,J212531.92$-$010745.9} were
obtained on ten nights during September and October 2005 with the
T\"ubingen 80\,cm and the G\"ottingen 50\,cm telescopes
(Fig.\,\ref{figlc}). 
\par
The light curve of SDSS\,J212531.92$-$010745.9 shows photometric
variability with a period of 6.9\,h and a peak-to-peak amplitude of
0.7\,mmag. The 6.9\,h periodicity is interpreted as the orbital period
of the system. A profile is obtained by folding the observations with
this period. It shows no eclipses (this constrains the inclination in
the light curve solution), a flat part, and periodic brightening that
is attributed to the light contribution by the irradiated side of the
cool companion.
\par
Collecting together spectroscopic parameters, reasonable assumptions,
photometric observations, and performing a fit to the folded profile
with the \linebreak
\texttt{nightfall} program, \citet{2006A&A...448L..25N} arrived at one
possible system solution as presented in Table~\ref{tablightcurve}.
A key parameter of interest, the mass of the primary, will however
have to await radial velocity measurements for an independent
determination.
\par
The light curve does not suggest the presence of pulsations in the
PG\,1159 star, but the constraints to exclude pulsations are weak:
The existing data exclude pulsations with amplitudes above 50\,mmag at
periods longer than about 8\,min.
\setcounter{figure}{0}
\begin{figure}[!t]
  \plotone{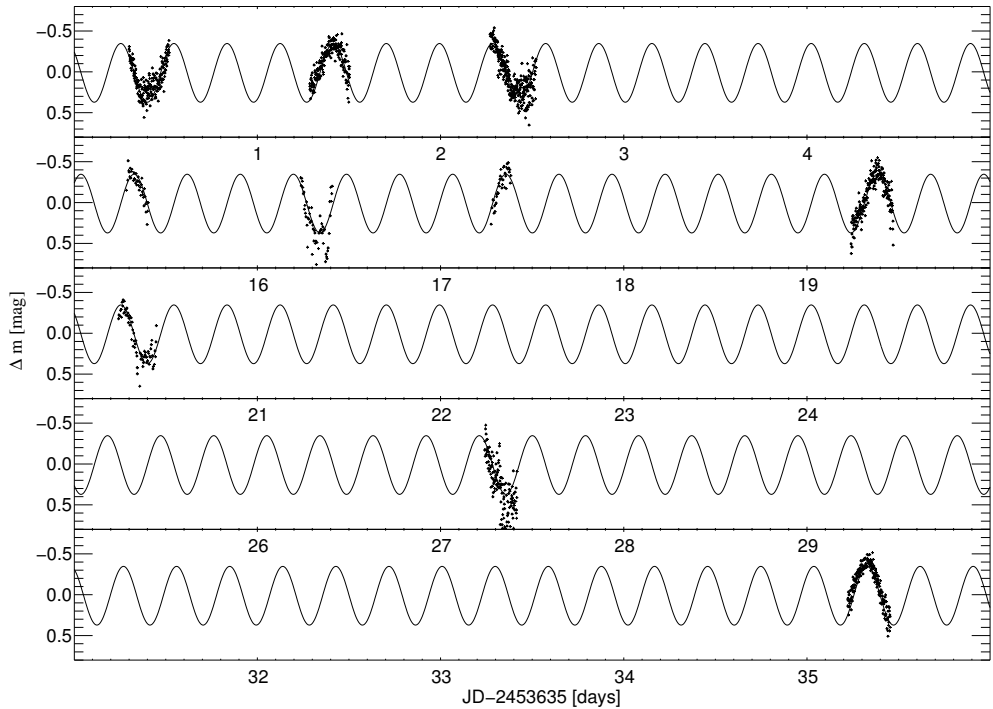}
  \caption{The photometric data from observing season 2005 (small
    crosses) can in the first instance be fit by a simple sine
    function (continuous line) with a 6.95616(33)\,h period 
    (adopted from Fig.\,3 in \citealt{2006A&A...448L..25N}).}
  \label{figlc}
\end{figure}
\hfill
\begin{table}[!t]
\caption{Resulting stellar and system parameters of SDSS\,J212531.92$-$010745.9,
  derived from comparison with NLTE model spectra (boldface), or
  assumed (normal font), and derived from photometric analysis ($^{\star}$) and
  a \texttt{nightfall} simulation (italic).}
\label{tablightcurve}
\smallskip
\begin{center}
{\small
\begin{tabular}{r@{}lr@{}c@{}lr@{}c@{}lr@{}c@{}l}
\tableline
\noalign{\smallskip}
    \multicolumn{2}{c}{parameter} &
    \multicolumn{3}{c}{PG\,1159 star}  &
    \multicolumn{3}{c}{companion} &
    \multicolumn{3}{c}{system} \\
    \noalign{\smallskip}
    \tableline
    \noalign{\smallskip}
    $T_{\rm eff}\,$&$\,[\rm K]$     &{\bf 90\,000}&{\bf$\pm$}&{\bf 20\,000} &3\,500&$\pm$&150&&&\\
    $T_{\rm eff,irr}\,$&$\, [\rm K]$&&&                   &{\it 8\,200}&&   &&&\\
    $\log\,(g$&$\,[\rm cm\,s^{-2}])$&{\bf 7.6}&{\bf $\pm$}&{\bf 0.5}&  &&   &&&\\
    $m\,$&$\,[\rm M_\odot]$         &0.6&&                &{\it 0.4}&{\it $\pm$}&{\it 0.1}&1.0&$\pm$&0.1\\
    $r\,$&$\,[\rm R_\odot]$         &0.1&&                &{\it 0.4}&{\it $\pm$}&{\it 0.1}&&&\\
    $P_{\rm orb}\,$&$\,[\rm h]$     &&&                   &&&                 & 6.95616$^{\star}$&$\pm$&0.00033\\
    $\Delta m\,$&$\,[\rm mag]$      &&&                   &&&                 & 0.354$^{\star}$&$\pm$&0.003\\
    $a\,$&$\,[\rm R_\odot]$         &&&                   &&&                 & 1.85&&\\
    $i\,$&$\,[^\circ]$              &&&                   &&& & {\it 70}&{\it $\pm$}&{\it 5}\\    
\noalign{\smallskip}
\tableline\
\end{tabular}
}
\end{center}
\end{table}
\clearpage
\subsection{Outlook}
Since the results cited above must be considered as preliminary, the
next step should be a full two-component analysis of orbital phase
resolved spectroscopy. These observations are in the queue, and will be
assisted by ongoing monitoring in the current 2006 observing season
which will allow us to much refine the ephemeris.
Ultimately, a precise mass for a PG\,1159 star derived with the help
of radial velocity measurements will allow to
check born-again scenarios of post-AGB evolutionary models.
\acknowledgements 
The authors thank M.~Schreiber, B.~G\"ansicke and S.~Dreizler for
pointing out to them the interesting SDSS spectrum of the object
discussed, and for their collaborative efforts in trying to understand
the system. They acknowledge the contributions of D.-J.~Kusterer,
T.~Stahn and many other observers, as well as the work of
S.~H\"ugelmeyer on the spectral analysis.
\par
S.~S.\ is particularly grateful for inspiring discussions with K.~Werner. 
\par
Travel to the 15th White dwarf workshop where this
contribution was presented has been subsidised by the Deutsche
Forschungsgemeinschaft (DFG) under grant number \mbox{KON~1082/2006
  SCHU~2249/2-1} to S.~S.


\begin{thebibliography}{}
\expandafter\ifx\csname natexlab\endcsname\relax\def\natexlab#1{#1}\fi

\bibitem[{{Bl\"ocker}(1995)}]{1995A&A...299..755B}
{Bl\"ocker}, T. 1995, \aap, 299, 755

\bibitem[{{Bond} \& {Ciardullo}(1999)}]{1999PASP..111..217B}
{Bond}, H.~E. \& {Ciardullo}, R. 1999, \pasp, 111, 217

\bibitem[{{Bond} {et~al.}(1996){Bond}, {Kawaler}, {Ciardullo}, {Stover},
  {Kuroda}, {Ishida}, {Ono}, {Tamura}, {Malasan}, {Yamasaki}, {Hashimoto},
  {Kambe}, {Takeuti}, {Kato}, {Kato}, {Chen}, {Leibowitz}, {Roth}, {Soffner},
  \& {Mitsch}}]{1996AJ....112.2699B}
{Bond}, H.~E., {Kawaler}, S.~D., {Ciardullo}, R., {et~al.} 1996, \aj, 112, 2699

%\bibitem[{Fu {et~al.}(2006)}{Fu}, {Vauclair}, \& {et~al.}]{fu:06}
%Fu, J.-N.,  Vauclair, G., \& et~al. 2006, \aap, submitted

\bibitem[{{Gonz{\'a}lez P{\'e}rez} {et~al.}(2006){Gonz{\'a}lez P{\'e}rez},
  {Solheim}, \& {Kamben}}]{2006A&A...454..527G}
{Gonz{\'a}lez P{\'e}rez}, J.~M., {Solheim}, J.-E., \& {Kamben}, R. 2006, \aap,
  454, 527

\bibitem[{H\"ugelmeyer {et~al.}(2007)H\"ugelmeyer, Dreizler, Werner,
  Krzesi\'nski, Nitta, \& Kleinman}]{huegelmeyer:07}
H\"ugelmeyer, S.~D., Dreizler, S., Werner, K., {et~al.} 2007, in ASP Conf.\ Ser., The 15th
  European Workshop on White Dwarfs, eds. R.~Napiwotzki \& M.~Burleigh (San Francisco: ASP), these proceedings

\bibitem[{Jahn {et~al.}(2007)Jahn, Reiff, Rauch, Werner, Kruk, \&
  Herwig}]{jahn:07}
Jahn, D., Reiff, E., Rauch, T., {et~al.} 2007, in ASP Conf.\ Ser., The 15th European Workshop on
  White Dwarfs, eds. R.~Napiwotzki \& M.~Burleigh (San Francisco: ASP), these
  proceedings

\bibitem[{{Kawaler} \& {Bradley}(1994)}]{kawaler:94}
{Kawaler}, S.~D. \& {Bradley}, P.~A. 1994, \apj, 427, 415

\bibitem[{{Kawaler} {et~al.}(1995){Kawaler}, {O'Brien}, {Clemens}, {Nather},
  {Winget}, {Watson}, {Yanagida}, {Dixson}, {Bradley}, {Wood}, {Sullivan},
  {Kleinman}, {Meistas}, {Leibowitz}, {Moskalik}, {Zola}, {Pajdosz},
  {Krzesinski}, {Solheim}, {Bruvold}, {O'Donoghue}, {Katz}, {Vauclair},
  {Dolez}, {Chevreton}, {Barstow}, {Kanaan}, {Kepler}, {Giovannini},
  {Provencal}, \& {Hansen}}]{kawaler:95}
{Kawaler}, S.~D., {O'Brien}, M.~S., {Clemens}, J.~C., {et~al.} 1995, \apj, 450,
  350

\bibitem[{{Kawaler} {et~al.}(2004){Kawaler}, {Potter}, {Vu{\v c}kovi{\'c}},
  {Dind}, {O'Toole}, {Clemens}, {O'Brien}, {Grauer}, {Nather}, {Moskalik},
  {Claver}, {Fontaine}, {Wesemael}, {Bergeron}, {Vauclair}, {Dolez},
  {Chevreton}, {Kleinman}, {Watson}, {Barstow}, {Sansom}, {Winget}, {Kepler},
  {Kanaan}, {Bradley}, {Dixson}, {Provencal}, \& {Bedding}}]{kawaler:04}
{Kawaler}, S.~D., {Potter}, E.~M., {Vu{\v c}kovi{\'c}}, M., {et~al.} 2004,
  \aap, 428, 969

\bibitem[{{Miller Bertolami} {et~al.}(2006){Miller Bertolami}, {Althaus},
  {Serenelli}, \& {Panei}}]{2006A&A...449..313M}
{Miller Bertolami}, M.~M., {Althaus}, L.~G., {Serenelli}, A.~M., \& {Panei},
  J.~A. 2006, \aap, 449, 313

\bibitem[{{Nagel} {et~al.}(2006){Nagel}, {Schuh}, {Kusterer}, {Stahn},
  {H{\"u}gelmeyer}, {Dreizler}, {G{\"a}nsicke}, \&
  {Schreiber}}]{2006A&A...448L..25N}
{Nagel}, T., {Schuh}, S., {Kusterer}, D.-J., {et~al.} 2006, \aap, 448, L25

\bibitem[{{Paunzen} {et~al.}(1998){Paunzen}, {K\"onig}, \&
  {Dreizler}}]{1998A&A...331..162P}
{Paunzen}, E., {K\"onig}, M., \& {Dreizler}, S. 1998, \aap, 331, 162

\bibitem[{Quirion(2007)}]{quirion:07}
Quirion, P.-O. 2007, in ASP Conf.\ Ser., The 15th European Workshop on White Dwarfs, eds.
  R.~Napiwotzki \& M.~Burleigh (San Francisco: ASP), these proceedings

\bibitem[{{Reed} {et~al.}(2000){Reed}, {Kawaler}, \&
  {O'Brien}}]{2000ApJ...545..429R}
{Reed}, M.~D., {Kawaler}, S.~D., \& {O'Brien}, M.~S. 2000, \apj, 545, 429

\bibitem[{Reiff {et~al.}(2007)Reiff, Jahn, Rauch, Werner, Kruk, \&
  Herwig}]{reiff:07}
Reiff, E., Jahn, D., Rauch, T., {et~al.} 2007, in ASP Conf.\ Ser., The 15th European Workshop on
  White Dwarfs, eds. R.~Napiwotzki \& M.~Burleigh (San Francisco: ASP), these
  proceedings

\bibitem[{{Sch\"onberner}(1983)}]{1983ApJ...272..708S}
{Sch\"onberner}, D. 1983, \apj, 272, 708

\bibitem[{Schreiber \& G\"ansicke(2007)}]{schreiber:07}
Schreiber, M. \& G\"ansicke, B. 2007, in ASP Conf.\ Ser., The 15th European Workshop on White
  Dwarfs, eds. R.~Napiwotzki \& M.~Burleigh (San Francisco: ASP), these
  proceedings

\bibitem[{Vauclair(2007)}]{vauclair:07}
Vauclair, G. 2007, in ASP Conf.\ Ser., The 15th European Workshop on White Dwarfs, eds.
  R.~Napiwotzki \& M.~Burleigh (San Francisco: ASP), these proceedings

\bibitem[{{Vauclair} {et~al.}(2002){Vauclair}, {Moskalik}, {Pfeiffer},
  {Chevreton}, {Dolez}, {Serre}, {Kleinman}, {Barstow}, {Sansom}, {Solheim},
  {Belmonte}, {Kawaler}, {Kepler}, {Kanaan}, {Giovannini}, {Winget}, {Watson},
  {Nather}, {Clemens}, {Provencal}, {Dixson}, {Yanagida}, {Nitta Kleinman},
  {Montgomery}, {Klumpe}, {Bruvold}, {O'Brien}, {Hansen}, {Grauer}, {Bradley},
  {Wood}, {Achilleos}, {Jiang}, {Fu}, {Marar}, {Ashoka}, {Me{\u i}stas},
  {Chernyshev}, {Mazeh}, {Leibowitz}, {Hemar}, {Krzesi{\'n}ski}, {Pajdosz}, \&
  {Zo{\l}a}}]{vauclair:02a}
{Vauclair}, G., {Moskalik}, P., {Pfeiffer}, B., {et~al.} 2002, \aap, 381, 122

\bibitem[{Werner {et~al.}(2007)Werner, Drake, Rauch, Schuh, \&
  Gautschy}]{werner:07}
Werner, K., Drake, J.~J., Rauch, T., Schuh, S., \& Gautschy, A. 2007,
  in ASP Conf.\ Ser., The 15th European Workshop on White Dwarfs,
  eds. R.~Napiwotzki \& M.~Burleigh (San Francisco: ASP), these proceedings

\bibitem[{{Werner} \& {Herwig}(2006)}]{2006PASP..118..183W}
{Werner}, K. \& {Herwig}, F. 2006, \pasp, 118, 183

\bibitem[{{Wesemael} {et~al.}(1985){Wesemael}, {Green}, \&
  {Liebert}}]{1985ApJS...58..379W}
{Wesemael}, F., {Green}, R.~F., \& {Liebert}, J. 1985, \apjs, 58, 379

\bibitem[{{Wood} \& {Faulkner}(1986)}]{1986ApJ...307..659W}
{Wood}, P.~R. \& {Faulkner}, D.~J. 1986, \apj, 307, 659

\end{thebibliography}
\end{document}